\documentclass[letterpaper, 10 pt, conference]{ieeeconf}  
\pdfoutput=1

\IEEEoverridecommandlockouts                              

\usepackage[ngerman, english]{babel}
\usepackage{graphicx} 
\usepackage{amsmath} 
\usepackage{siunitx}
\usepackage{cleveref}
\crefname{figure}{Fig.}{Fig.}
\crefname{equation}{}{}
\Crefname{equation}{Equation}{Equations}
\usepackage{booktabs}

\usepackage{hologo}
\usepackage{listings}
\title{\LARGE \bf
On Assumptions with Respect to Occlusions in Urban Environments \\for Automated Vehicle Speed Decisions*
}

\author{Robert Graubohm$^{1}$, Nayel Fabian Salem$^{1}$, Marcus Nolte$^{1}$, and Markus Maurer$^{1}$
\thanks{*This research is accomplished within the projects ``UNICAR\emph{agil}'' (FKZ~16EMO0285) and ``AUTOtechagil'' (FKZ~01IS22088R). We acknowledge the financial support for both projects by the Federal Ministry of Education and Research of Germany (BMBF).\newline This work has partially been funded by the Daimler and Benz Foundation.}
\thanks{$^{1}$All authors are with the Institute of Control Engineering at Technische Universit\"at Braunschweig, 38106 Braunschweig, Germany
        {\tt\small \{graubohm,salem,nolte,maurer\}@ifr.ing.tu-bs.de}}%
}

\begin{document}

\twocolumn[
\begin{@twocolumnfalse}
			\Huge {IEEE copyright notice} \\ \\
	\large {\copyright\ 2023 IEEE. Personal use of this material is permitted. Permission from IEEE must be obtained for all other uses, in any current or future media, including reprinting/republishing this material for advertising or promotional purposes, creating new collective works, for resale or redistribution to servers or lists, or reuse of any copyrighted component of this work in other works.} \\ \\
	
	{\Large Published in \emph{2023 IEEE 26th International Conference on Intelligent Transportation Systems (ITSC)}, Bilbao, Spain, September 24-28, 2023.} \\ \\ 
{\Large DOI: \href{https://doi.org/10.1109/ITSC57777.2023.10422457}{10.1109/ITSC57777.2023.10422457}} \\ \\

Cite as:
\vspace{0.1cm}

\noindent\fbox{%
    \parbox{\textwidth}{%
        R.~Graubohm, N.~F. Salem, M.~Nolte, and M.~Maurer, ``On {Assumptions} with {Respect} to {Occlusions} in {Urban} {Environments} for {Automated} {Vehicle} {Speed} {Decisions},''
  in \emph{2023 {IEEE} {Int.} {Conf.} {Intelligent} {Transportation} {Systems}},\hskip 1em plus
  0.5em minus 0.4em\relax  Bilbao, Spain, 2023, pp. 738--745, doi: {10.1109/ITSC57777.2023.10422457}.
    }%
}
\vspace{2cm}

\end{@twocolumnfalse}
]

\noindent\begin{minipage}{\textwidth}

\hologo{BibTeX}:
\footnotesize
\begin{lstlisting}[frame=single]
@inproceedings{graubohm_assumptions_2023,
  author={{Graubohm}, Robert and {Salem}, Nayel Fabian and {Nolte}, Marcus and {Maurer}, Markus},
  booktitle={2023 {IEEE} 26th {International} {Conference} on {Intelligent} {Transportation} {Systems} {({ITSC})}},
  title={On {Assumptions} with {Respect} to {Occlusions} in {Urban} {Environments} for {Automated} {Vehicle} {Speed} {Decisions}},
  address={Bilbao, Spain},
  year={2023},
  pages={738--745},
  doi={10.1109/ITSC57777.2023.10422457},
  publisher={IEEE}
}
\end{lstlisting}
\end{minipage}

\maketitle
\thispagestyle{empty}
\pagestyle{empty}

\begin{abstract}

Automated driving systems are subject to various kinds of uncertainty during design, development, and operation. These kinds of uncertainty lead to an inherent risk of the technology that can be mitigated, but never fully eliminated.
Situations involving obscured traffic participants have become popular examples in the field to illustrate a subset of these uncertainties that developers must deal with during system design and implementation. 
In this paper, we describe necessary assumptions for a speed choice in a situation in which an ego-vehicle passes parked vehicles that generate occluded areas where a human intending to cross the road could be obscured. We develop a calculation formula for a dynamic speed limit that mitigates the collision risk in this situation, and investigate the resulting speed profiles in simulation based on example assumptions. This paper has two main results: First, we show that even without worst-case assumptions, dramatically reduced speeds would be driven to avoid collisions. Second, we highlight that design decisions regarding occlusion treatment are directly related to the risk that automated vehicles pose to pedestrians in urban environments. In this respect, we conclude that there needs to be a broader discussion about acceptable assumptions.

\end{abstract}

\section{INTRODUCTION}
\label{sec:intro}

There is no question that the complete avoidance of all possible collisions in urban traffic is illusory. There is always risk associated with the operation of a vehicle in the traffic system. On the one hand, there is inherent risk in today's traffic system that cannot be eliminated~\cite{nolte_supporting_2020}. On the other hand, as with any technical system, there is certain risk associated with technology itself, and the concept of product safety is primarily concerned with whether this risk is acceptable~\cite{salem_risk_2023}. ISO/IEC~GUIDE~51 notes: ``Some level of risk is inherent in products or systems''~\cite[Clause~4.1]{international_organization_for_standardization_isoiec_2014}. It is important to consider this fact when discussing the safety potential of future road traffic automation. In particular, the contribution of automated road vehicles to the EU's \emph{Vision Zero} of ``as close as possible to zero fatalities in road transport by 2050''~\cite[p.~4]{european_climate_infrastructure_and_environment_executive_agency_eu_2022} needs to be critically examined.

When developing complex vehicular systems, such as automated driving systems, engineers are required to analyze residual risk (both inherent and resulting from design decisions) and to evaluate it with respect to risk acceptance criteria~\cite[Clause~6]{international_organization_for_standardization_iso_2022}. These criteria are expected to reflect societal acceptance of the technology despite inherent risk. Informed public acceptance requires open risk communication, which includes disclosing the results of risk analyses and arguing for the appropriateness of design decisions. To date, however, these risk acceptance criteria for automated driving applications have not been generally agreed upon in the course of the necessary public debate~\cite{resnik_precautionary_2023,maurer_wirtschaft_2005}.

As an illustrative example of challenging design decisions, in this paper we use a situation that is often associated with unavoidable collisions: A pedestrian suddenly emerges from an occluded area and steps in front of an approaching vehicle~\cite{maurer_hochautomatisiertes_2018,mobileye_implementing_2018,nister_introduction_2019,wang_potential_2022,jeong_collision_2020}. Accidents involving pedestrians, who are among the vulnerable road users, represent one of the worst cases in traffic and must be strictly avoided by road vehicles, as life-threatening injuries can occur even at relatively low speeds. Occlusions represent the particular case where there is epistemic uncertainty about the existence and future actions of pedestrians (cf.~\cite{nolte_representing_2018}). Significant speed reduction can essentially make many such encounters controllable for a vehicle guidance system, through timely response and shorter stopping distances (as represented, e.g., in Mobileye's RSS framework, cf. \Cref{sec:relatedwork}). However, automated vehicles that drive at lower speeds can negatively impact traffic flow and acceptance by other road users. In this paper, we present a calculation formula for maximum travel speeds while passing occluded areas, with the goal of still allowing stopping before a collision. However, we find that assumptions must be made to determine which situations are still controllable and which situations end in unavoidable collisions and hence contribute to the system's residual risk. IEEE~Std~2846~\cite{intelligent_transportation_systems_committee_2846-2022_2022} also emphasizes the importance of considering assumptions about the behavior of other road users in safety-related models for automated driving systems. Making these decisions individually is beyond the competence of the developers of such functions and should rather be based on a broad public consensus.

Consequently, an important contribution of our speed limit modeling is that assumptions deciding what behavior of automated vehicles in urban environments is considered \emph{safe} are explicitly specified as part of their design and safety argument. We introduce examples of assumptions to analyze the effects on absolute travel speeds, accelerations, etc., and show the side effects of such distinctly cautious vehicle behavior. To present the scope of this paper, in \Cref{sec:situation} we describe the considered traffic situation, including some observations on its regulatory treatment, before introducing related work in \Cref{sec:relatedwork}. We then present our model in detail in \Cref{sec:model}, evaluate it in \Cref{sec:eval}, and finally discuss our results in \Cref{sec:discussion}.

\section{BACKGROUND}
\label{sec:situation}
\subsection{Considered Traffic Situation}

\begin{figure}[t]
\centering
\includegraphics[scale=0.125]{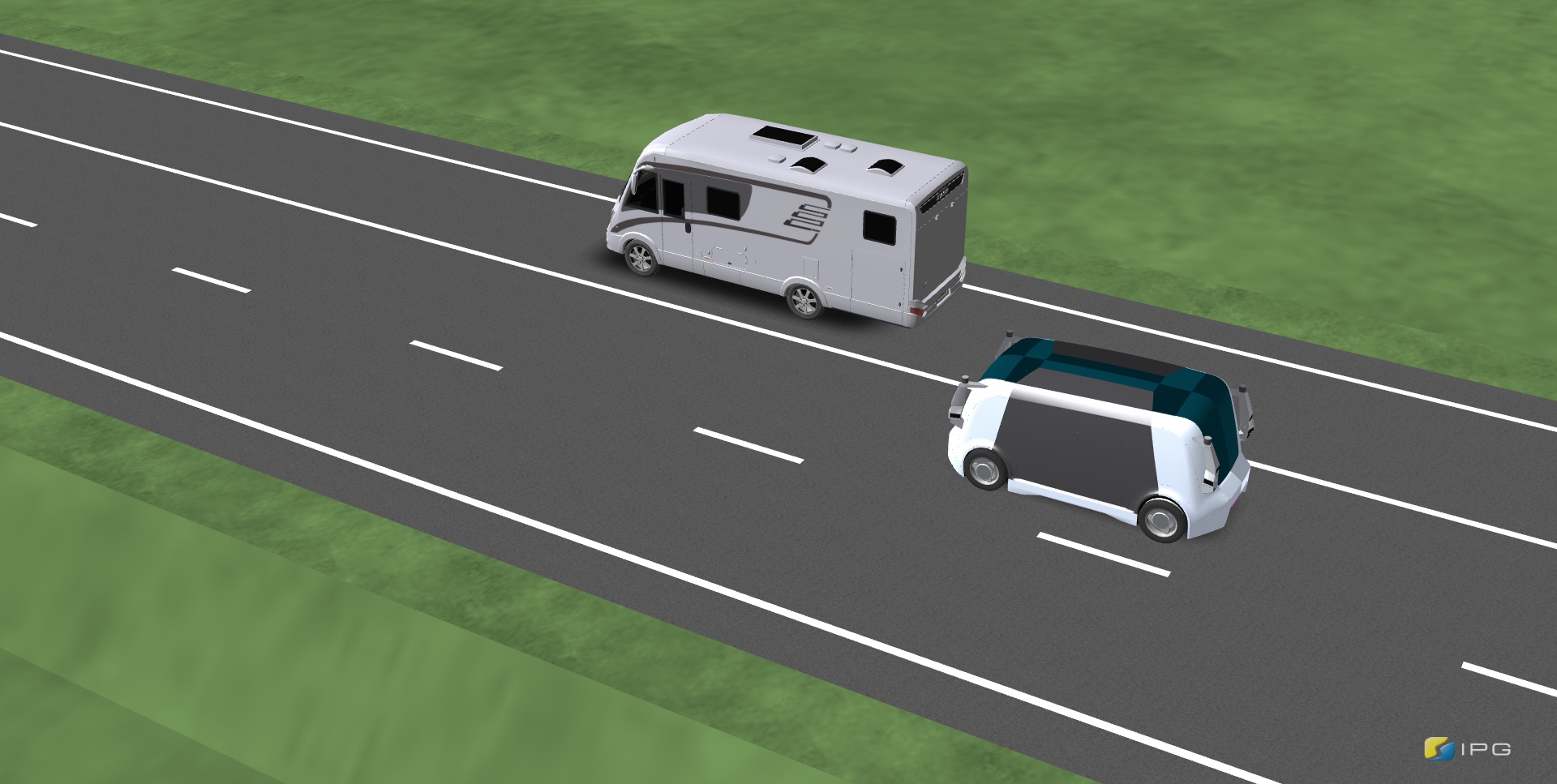}
\caption{Scenario examined of an automated vehicle passing a large parked vehicle}
\label{fig:IniSituation}
\end{figure}

In this paper, we focus entirely on a situation in which an (automated) ego-vehicle passes one or more larger parked vehicles relatively closely in an urban environment. In \cref{fig:IniSituation}, we illustrate such a situation in the simulation environment CarMaker by IPG Automotive\footnote{https://ipg-automotive.com/en/products-solutions/software/carmaker/}, which we use for evaluation purposes in \Cref{sec:eval}. While passing, every vehicle parked next to the ego-lane causes a potentially relevant occluded space. The shape of the occluded space depends on the number and positioning of the ego-vehicle's environment sensors, as well as on the geometry and type of the parked vehicle being passed. The potential to completely obscure a human in this situation exists and is a regular occurrence in today's road traffic. Children are of particular concern. Their non-detectable presence is possible even in relatively small occluded areas, which increases the incidence of occlusions that could theoretically obscure a child. Additionally, young children's lack of knowledge of traffic rules may cause them to carelessly step into the road despite the passing traffic having the right of way.

\subsection{Tests Regarding Children Emerging from Occlusions}

Mandatory or voluntary standardized tests of vehicles are an indicator of the expected behavior and the capabilities to react in the traffic situation we described above. For example, test scenarios that can result from the initial situation are already part of the evaluation of automatic emergency braking systems with vulnerable road user protection. In the Car-to-Pedestrian Nearside Child \qty{50}{\percent} (CPNC-50) scenario of the Euro NCAP~\cite{european_new_car_assessment_programme_test_2021}, a child pedestrian target (as specified in ISO~19206-2:2018) emerges from an occlusion caused by a large parked vehicle at a speed of \qty{5}{\kilo\metre\per\hour} while the vehicle under test is approaching. The perpendicular child scenario performed by IIHS~\cite{insurance_institute_for_highway_safety_pedestrian_2022} for such systems is highly similar. The test result for vehicles with automatic emergency braking is based on their potential to prevent collisions given different driving speeds.

The potential to use a scenario like the CPNC-50 alongside other Euro NCAP scenarios to evaluate automated driving systems has already been highlighted by OCIA in 2019~\cite[p.~40]{international_organization_of_motor_vehicle_manufacturers_future_2019}. However, the authors also point out the difficulty that an automated vehicle in this situation would not necessarily follow a set speed. The movement of the child pedestrian target would need to be synchronized with the actual travel speed of the vehicle under test in order for a collision to occur at all if the vehicle does not react to the pedestrian.
Regulation (EU)~2022/1426, which establishes procedures and technical specifications for the type-approval of the automated driving system of fully automated vehicles, already requires a number of collision avoidance tests to be performed by an automated vehicle, including scenarios with pedestrians~\cite[Annex~3, Part~3]{european_commission_directorate-general_for_internal_market_industry_entrepreneurship_and_smes_commission_2022}. However, it does not yet include a test scenario involving a child emerging from an occlusion.

Overall, it should be noted that the tests mentioned focus on testing the ability of technical systems to react promptly and correctly in the event of an imminent collision. They do not provide any specific guidance on proactive speed reduction when facing occlusions from which a child might emerge.
In addition, it should be emphasized that the standardized tests operate with limited pedestrian speeds. Regulation (EU)~2022/1426 also sets a pedestrian speed of \qty{5}{\kilo\metre\per\hour} for collision avoidance tests of automated driving systems~\cite[Annex~3, Part~3, 8.5.1]{european_commission_directorate-general_for_internal_market_industry_entrepreneurship_and_smes_commission_2022}.
Given that even young children can reach twice this speed when running~\cite{schepens_mechanics_1998}, the generalizability of results from such tests in an open-world context is significantly limited. Consequently, our model presented in \Cref{sec:model} allows for smaller lateral distances, different pedestrian positions, and higher pedestrian speeds than those assumed in standardized tests.

\subsection{Regulations On Assumptions with Respect to Occlusions}

While currently each driver has to defend their individual speed choice in case of an accident, the behavior of automated driving systems is defined by developers. As a result, the developers of automated vehicles will be responsible for the speed decisions of entire fleets (with important legal liability issues remaining unresolved, cf.~\cite{european_commission_directorate_general_for_research_and_innovation_ethics_2020}). There is no question that, primarily, posted or general route-specific speed limits must be observed. However, the appropriate driving speed may be further limited by the specific traffic situation. In this context, permitting vehicles to travel at relatively high speeds, even when pedestrians are or may be in close proximity, is often enabled by the fact that pedestrians are also presumed to obey traffic laws to a certain extent. Drivers do not have to assume that pedestrians on sidewalks will spontaneously jaywalk at running speed. However, in Germany, for example, this \emph{principle of trust} is limited to adults who do not show clear signs of impaired judgment~\cite{feldmanis_principle_2019,frey_psychologische_2018}. Yet today's actual urban road traffic shows that (even prudent) drivers usually do not drive extremely slowly when passing occluded areas to account for the possibility of an obscured child suddenly stepping in front of their car.

Although automated vehicles are clearly expected to operate with a lower risk of collision than human drivers (often expressed by the term ``positive risk balance''~\cite{international_organization_for_standardization_isotr_2020}), there is limited regulatory guidance on the assumptions to be made for speed decisions. Regulation (EU)~2022/1426 requires an automated driving system to demonstrate ``risk minimising behaviour when critical situations could become imminent, e.g. with unobstructed and obstructed vulnerable road users (pedestrians, cyclist, etc.)''~\cite[Annex~2, Part~3, 1.1.3]{european_commission_directorate-general_for_internal_market_industry_entrepreneurship_and_smes_commission_2022}, without further characterizing this behavior. In fact, the EU regulation defines that collisions have to be avoided at least as long as a certain minimum time-to-collision is not undercut by the crossing of a pedestrian~\cite[Annex~3, Part~1, 1.4.2]{european_commission_directorate-general_for_internal_market_industry_entrepreneurship_and_smes_commission_2022}. However, this minimum time-to-collision for collision avoidance is based on a speed-dependent equation. As a result, the threshold increases as the speed of the automated vehicle increases. In addition to generally requiring collision avoidance with visible pedestrians in urban and rural environments, Regulation (EU)~2022/1426 demands a speed reduction of at least \qty{20}{\kilo\metre\per\hour} before collision in the case of pedestrians crossing out of an occlusion~\cite[Annex~3, Part~1, 1.4.3]{european_commission_directorate-general_for_internal_market_industry_entrepreneurship_and_smes_commission_2022}, which, in some scenarios, seems physically impossible to us given arbitrarily short reaction and deceleration distances.

Even more limited than the EU requirements, the US Department of Transportation's report Automated Vehicles 4.0 only points to the potential of automated vehicles to improve the physical safety of pedestrians~\cite[p.~4]{national_science__technology_council_ensuring_2020}. Furthermore, the US Federal Automated Vehicles Policy specifies that automated vehicles must ``yield to pedestrians'' and ``provide safe distance''~\cite[p.~29]{national_highway_traffic_safety_administration._federal_2016}. In summary, regulatory documents generally indicate what behavior characteristics are expected to be implemented in automated vehicles and their systems, but they do not provide concrete guidance on context-specific driving speeds.

\section{RELATED WORK}
\label{sec:relatedwork}

Two types of related work can be highlighted regarding assumptions for speed decisions of automated vehicles with respect to occlusions. On the one hand, there are publications on formal rules to explicitly describe safe behavior. On the other hand, there are approaches that consider the uncertainty due to occluded areas in algorithms for vehicle guidance.

An example for the representation of safe vehicle behavior in formal rules is the Responsibility-Sensitive Safety (RSS) assurance framework for validating the safety of automated driving functions~\cite{shalev-shwartz_formal_2017}.
The primary focus of RSS, however, is to exclude responsibility, not to argue for the absence of unreasonable risk. An example of a mathematical description within RSS is the formula for ``safe longitudinal distances'' when following another vehicle. The aspect of interaction with vulnerable road users is also discussed in the publications on the RSS approach. Yet, for the scenario we analyze in this paper (see \Cref{sec:situation}), the explanations in the RSS documentation seem to be contradictory. On the one hand, the approach positions itself in such a way that the speed adjustment under the assumption that a person could emerge from any occlusion causes a behavior that is ``too defensive''~\cite[p.~14]{mobileye_implementing_2018} or ``over defensive, non natural driving''~\cite[p.~24]{shalev-shwartz_formal_2017}. This seems to imply that there would not be extreme restrictions on driving speed due to the risk of pedestrian accidents around occlusions. On the other hand, the framework specifies two circumstances as criteria that must be met for the automated vehicle to be considered~-- according to RSS~-- not responsible for an accident with a pedestrian coming out of an occlusion~\cite[p.~16]{mobileye_implementing_2018}. The first criterion is that the vehicle stops accelerating from the moment the pedestrian can first be seen by the vehicle. The second criterion is that the average speed of the vehicle in the time interval between the moment of emergence and the accident is lower than the speed of the pedestrian.

To meet the second criterion ``the average velocity of the vehicle from the exposure time until the collision time must be smaller than the average velocity of the pedestrian at the same time interval''~\cite[p.~16]{mobileye_implementing_2018}, the vehicle must initially drive quite slowly, considering the usual walking speed of pedestrians. As a result, RSS may require speeds of less than \qty{10}{\kilo\metre\per\hour} near occluded areas even before a pedestrian appears. So in summary, RSS suggests~-- as we do in this paper~-- that significant speed reductions would be needed around occlusions to substantially mitigate the risk of collision. However, the RSS authors do not disclose what assumptions went into their criteria or how they assess the residual risk.

From an implementation point of view, numerous other works and projects also address explicit treatment of occluded areas. When introducing their Safety Force Field, Nistér et al.~\cite{nister_introduction_2019} formulate that they want to handle all occlusions by assuming relevant actors in them and consequently adjusting speed and lateral distances. However, implementation details, including the assumptions incorporated in the Safety Force Field, are not provided by Nistér et al.

Wang et al.~\cite{wang_potential_2022} develop a dynamic risk assessment for handling potentially occluded pedestrians based on the Bayesian theory and incorporated into the motion planning. In their approach, based on a quantified potential risk, the authors determine a desired vehicle speed in relation to the speed limit of the current road section. Naumann et al.~\cite{naumann_safe_2019} focus on intersections and consider the probability of other vehicles moving in occluded areas when determining a speed profile. Yu et al.~\cite{yu_occlusion-aware_2019} also base their motion planning on a probabilistic risk assessment that specifically considers occlusions at intersections. Bouton et al.~\cite{bouton_scalable_2018} present a POMDP (partially observable Markov decision process) approach for decision making in the presence of occlusions, using occluded pedestrians at a crosswalk as one of their scenarios. Schratter et al.~\cite{schratter_pedestrian_2019} also formulate the problem of uncertain pedestrian locations in occlusions as a POMDP. Using scenarios from the Euro NCAP tests, they evaluate their approach with an autonomous emergency braking system.

Likewise, Jeong et al.~\cite{jeong_collision_2020} motivate their collision preventive velocity planning with occluded regions from which pedestrians might emerge. The authors' objectives and calculation formulas are in parts similar to our application example, but they assume very high potential vehicle decelerations in case of emergency braking, and their simulation focuses on a specific distance between the vehicle and a suddenly appearing pedestrian. Furthermore, the effects of their approach on the speed profile when approaching and passing occlusions without an emerging relevant object are not presented by Jeong et al.

\section{BASIC MODEL FOR SPEED DECISIONS WITH RESPECT TO OCCLUSIONS}
\label{sec:model}

In this paper, we demonstrate explicit occlusion treatment using a calculation formula for a dynamic speed limit. We assume an automated driving system designed to adhere to the calculated speed limit without compromise. This explicit rule makes the implementation of the occlusion treatment traceable and individually evaluable, which strengthens its potential role in the system-wide safety argument of the automated vehicle. In order to calculate a dynamic speed limit based on assumptions and evaluate its effect in simulation, we first have to represent the driving situation in a basic model.
   
In essence, our model is intended to describe a pedestrian stepping out of a sufficiently large occlusion onto the road at any point in time. Thus, we consider any occlusion that allows the undetected presence of a relevant object to be occupied by a person (possibly a child) with the intent to cross.
Using this model, we want to calculate a speed limit that enables collision avoidance, provided a number of assumptions are met. The main assumptions to be made here are braking capabilities, reaction times, pedestrian speed and geometry, and the capability of a perception system to detect occlusions. We introduce example assumptions for these parameters for evaluation purposes in \Cref{ssec:assump}.

In general, we assume that for an automated driving system, occlusions and their concrete geometric dimensions are part of the information used for motion planning. Based on the location and geometry of occlusions, we can model whether and how close to the lane a human could be obscured. 
Ultimately, we base a mathematical formulation of a speed supremum on the occlusion model, where observing the calculated speed supremum allows for collision avoidance unless the assumptions made are not met.

\begin{figure}[tbh]
      \centering
      \includegraphics[scale=0.7]{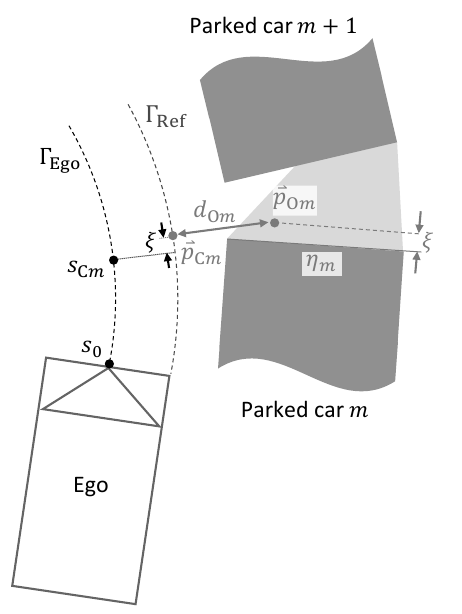}
      \caption{Ego-vehicle passes a parked car $m$ while following path $\mathrm{\Gamma}_\mathrm{Ego}$. From the ego-vehicle's perspective, there is an occluded area between parked cars $m$ and $m+1$ (indicated in the graphic). At $\vec{p}_{\mathrm{O}m}$ an obscured relevant object could be present.}
      \label{fig:Scene1}
   \end{figure}

\cref{fig:Scene1} shows the basic situation for the development of our calculation formula. The ego-vehicle follows the path $\mathrm{\Gamma}_\mathrm{Ego}$ and is located at $s_0$ at the given time $t_0$. Next to the ego-path is a parked vehicle $m$, which due to its geometry creates an occluded area next to the driving corridor for the ego-vehicle's environment sensors.

Among all locations where an obscured relevant object (e.g. child) could be present in the occlusion, $\vec{p}_{\mathrm{O}m}$ has the shortest distance to the driving corridor of the ego-vehicle. The path $\mathrm{\Gamma}_\mathrm{Ref}$ describes the boundary of the driving corridor towards the occlusion. We use the front line of vehicle $m$, $\mathrm{\eta}_m$, as a reference for the occluded area. $\vec{p}_{\mathrm{O}m}$ and $\vec{p}_{\mathrm{C}m}$ describe locations of a human, with respect to the front center of the human body (foremost point in the midline). Accordingly, there is a necessary minimum distance $\mathrm{\xi}$ to the object edge $\mathrm{\eta}_m$, which corresponds to half of an assumed human width. To describe the collision point, $\mathrm{\xi}$ has to be considered along $\mathrm{\Gamma}_\mathrm{Ref}$ as the difference between the ego-vehicle front ($s_{\mathrm{C}m}$ along $\mathrm{\Gamma}_\mathrm{Ego}$) and the pedestrian ($\vec{p}_{\mathrm{C}m}$). The distance $d_{\mathrm{O}m}$ traveled by a person potentially emerging from the occlusion is assumed to be the shortest distance between $\mathrm{\Gamma}_\mathrm{Ref}$ and $\vec{p}_{\mathrm{O}m}$. The pessimistic simplification of assuming that the obscured human could take a direct path to the driving corridor~-- regardless of any geometric minimum distances to the bounding box representing a parked vehicle~-- seems reasonable to us, since many vehicles are strongly rounded at the corners.

Based on this model for describing an occlusion and a potential conflict area, a collision avoiding speed can now be determined for the ego-vehicle. To quantify such a speed limit, the main elements to consider are reaction time and stopping distance.
When calculating a speed limit, our model is limited to the scenario in which the ego-vehicle collides frontally (in the direction of travel) with a pedestrian. An alternative collision scenario would be the previously obscured person running into the side of the ego-vehicle. The severity of such an accident can be affected by the shape and speed of the ego-vehicle, which should be considered when deciding on driving speeds in urban environments. In general, however, the collision cannot be avoided by altering the driving speed: Even if the ego-vehicle is stopped, a pedestrian could run into its side.

Thus, our first criterion for specifying a speed requirement is to check whether it is possible for a pedestrian to step in front of the ego-vehicle before it has passed the conflict area with its front line. For this check we use an assumed maximum pedestrian speed of $v_\mathrm{O}$ and the smallest possible distance $d_{\mathrm{O}m}$:
\begin{equation}
\label{eqn:d_cond}
\frac{d_{\mathrm{O}m}}{v_\mathrm{O}} \le t_{\mathrm{C}m} \text{, with } s(t_{\mathrm{C}m})=s_{\mathrm{C}m}\text{.}
\end{equation}
If \cref{eqn:d_cond} is not satisfied, the pedestrian can no longer step in front of the ego-vehicle. However, in situations where \cref{eqn:d_cond} is satisfied, collision avoidance would require that the stopping distance of the ego-vehicle during emergency braking with an assumed average maximum deceleration $a_\mathrm{max}$ after a reaction time $t_\mathrm{R}$ is smaller than the remaining distance to a potential collision point:
\begin{equation}
\label{eqn:d_supremum}
\frac{v_\mathrm{Ego}^2(t=t_0+t_\mathrm{R})}{2|a_\mathrm{max}|}+\int_{t_0}^{t_0+t_\mathrm{R}}v_\mathrm{Ego}(t)\mathrm{d}t < s_{\mathrm{C}m}-s_0\text{.}
\end{equation}
The highest possible $v_\mathrm{Ego}$ that satisfies \cref{eqn:d_supremum} is a speed supremum that in some situations can be far below the official speed limits. The consideration of a time-variant driving speed during a reaction time in \cref{eqn:d_supremum} limits the possible concretization of the speed requirement. An approximation of the future speed profile~-- e.g., based on the trajectory planning~-- is imaginable in the context of automated driving, but it creates a high degree of complexity and dependency.

For the application example in this paper, we therefore choose a strong simplification of the calculation formula by assuming a constant driving speed over the reaction time. Furthermore, we define the remaining distance until reaching the assumed collision point with a person emerging from the occlusion caused by parked vehicle $m$ as \mbox{$\mathrm{\Delta}s_m=s_{\mathrm{C}m}-s_0$}. As explained earlier, we consider a given occlusion as relevant for the speed decision of the ego-vehicle only until reaching the virtual collision point. Thus, the condition for applying a speed limit due to a certain occlusion, simplified by assuming a constant driving speed over the reaction time, is:
\begin{equation}
\label{eqn:c_cond}
\frac{d_{\mathrm{O}m}}{v_\mathrm{O}} \le \frac{\mathrm{\Delta}s_m}{v_\mathrm{Ego}}\land s_{\mathrm{C}m}>s_0\text{, }v_\mathrm{Ego}>0\text{.}
\end{equation}
Additionally, under the assumption of a constant speed in the reaction time, a direct calculation formula for a speed supremum $v_\mathrm{sup}$ can now be derived from \cref{eqn:d_supremum} by solving our quadratic equation for $v_\mathrm{Ego}$:
\begin{equation}
\label{eqn:ex_supremum}
v_\mathrm{Ego} < \underbrace{\sqrt{t_\mathrm{R}^2|a_\mathrm{max}|^2+2|a_\mathrm{max}|\mathrm{\Delta}s_m} -t_\mathrm{R}|a_\mathrm{max}|}_{%
\let\scriptstyle\textstyle
\substack{v_\mathrm{sup}}}\text{.}
\end{equation}

\begin{figure}[t]
      \centering
      \includegraphics[scale=0.7]{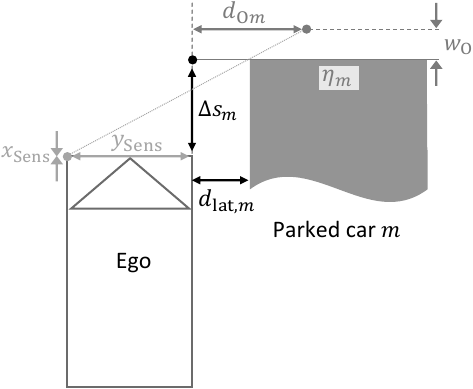}
      \caption{Simplified relations when passing a large vehicle parked next to the lane on a straight route}
      \label{fig:Scene2}
   \end{figure}

In addition, if we assume a straight lane, as shown in \cref{fig:IniSituation} instead of a curved lane, we can simplify the occlusion model considerably.
The recreational vehicle depicted represents a challenging occlusion: Vehicle sensors for environment perception, just like human drivers, cannot see through or over the parked vehicle in the case shown.
Therefore, the presence of an obscured person in the occluded area who could suddenly step onto the road cannot be ruled out until sensors can see past the front of the recreational vehicle. \cref{fig:Scene2} illustrates the described situation for an environment sensor positioned at the left front corner of the ego-vehicle: The UNICAR\emph{agil} vehicles~\cite{woopen_unicaragil_2018} have sensor modules with \ang{270} detection range at each vehicle corner (see \cref{fig:IniSituation}).

In the simplified case illustrated, the collision point of the ego-vehicle with an emerging pedestrian can be constructed as the intersection of the front line of the parked vehicle with the driving corridor of the ego-vehicle. For the closest complete obscuring of a person by the occlusion, an assumed total human width $w_\mathrm{O}=2\mathrm{\xi}$ is decisive. Without simulation of an environment perception, the minimum distance of an obscured human that could emerge from the occlusion can be described by:
\begin{equation}
\label{eqn:ex_distOgen}
d_{\mathrm{O}m}=d_{\mathrm{lat},m}+\frac{d_{\mathrm{lat},m}+y_\mathrm{Sens}}{\mathrm{\Delta}s_m-x_\mathrm{Sens}}w_\mathrm{O}\text{.}
\end{equation}
Here, $x_\mathrm{Sens}$ and $y_\mathrm{Sens}$ describe the distance of an effective sensor origin of the environment perception starting from the front right corner of the ego-vehicle (i.e., the contour point closest to the occlusion). From \cref{eqn:ex_distOgen} it is apparent that $d_{\mathrm{O}m}$, and thus the satisfaction of the condition for speed limitation expressed in \cref{eqn:c_cond}, is largely determined by the lateral distance of the ego-vehicle from the parked vehicle $d_{\mathrm{lat},m}$. Nolte et al.~\cite{nolte_representing_2018} show very clearly that maximizing the lateral distance in path planning is the most important measure for dealing with uncertainties due to external factors such as occlusions. Consequently, in our calculation formula, large lateral distances from occluded areas can also significantly reduce the effect of the calculated speed limit. In a situation with substantial lateral distances to occlusions, the dynamic speed limitation formula will either have no effect at all or only a brief effect with an insignificant speed reduction. Evaluating the effects of the calculated driving speed limit, which we will do in the next section, is therefore only meaningful for road sections where maintaining sufficient lateral distances from occluded areas is not feasible due to infrastructure or traffic conditions.

\section{EVALUATION BASED ON\\EXAMPLE ASSUMPTIONS}
\label{sec:eval}

To evaluate our model in simulation, we must first make assumptions about various parameters. One approach is to assume extreme values for each parameter to construct edge cases. For example, Mobileye assumes speeds at the limit of what is physically possible for a human being for the pedestrian emerging from an occlusion, noting that this ``yields behavior that is too defensive''~\cite[p.~14]{mobileye_implementing_2018}, and further commenting that:
\begin{quote}
``It follows that we should drive slower than \qty{1}{\metre\per\second} in this scenario, even if we drive at a lateral distance of, say, \qty{5}{\metre}, from the parked cars. This is too defensive and does not reflect normal human behavior.''~\cite[p.~14]{mobileye_implementing_2018}
\end{quote}

We can reproduce this finding in our model by assuming very high pedestrian speeds.
Moreover, if we try to make worst-case assumptions for all relevant parameters (very slow reaction, deceleration on black ice, and pedestrians running like sprinters out of occlusions), we can easily create a situation where a simulation is not meaningful because the planning of vehicle trajectories would involve minimal speeds that do not satisfy any mobility needs.
The result of such extremely pessimistic assumptions is shown by the dashed line in \cref{fig:SpeedLimitProfile}. Already \qty{50}{\metre} before reaching the parked vehicle, the calculated speed limit is significantly less than half of the posted speed limit. As the ego-vehicle actually passes the parked vehicle, the calculated speed limit drops to less than \qty{0.01}{\metre\per\second}.

\subsection{Example Assumptions for Speed Limit Calculation}
\label{ssec:assump}
As a complement to extremely pessimistic considerations, in this section we identify parameter values for our simulation that are based on established values from the literature. By doing so, we can demonstrate that even without worst-case assumptions, explicit occlusion treatment results in significant speed limitations that differ considerably from today's traffic patterns.

First, we use a pedestrian walking speed of \qty{1.6}{\metre\per\second} (following ISO~13855~\cite{international_organization_for_standardization_iso_2010}) for our example, which is slightly higher than the speed of the child pedestrian target in the Euro NCAP test. Second, to calculate the speed supremum for the ego-vehicle, we assume a reaction time $t_\mathrm{R}=\qty{1}{\second}$. While the time required to classify and track newly detected road users may be \qty{0.4}{\second} and less, depending on the perception system~\cite{rieken_lidar-based_2020}, a reaction time must also represent the entire duration until an actual system response (i.e., actuation of the dynamics systems) occurs. And a large reaction time can also be used to express that a dynamic speed limit is never instantaneously implemented by the ego-vehicle. Thus, the real speed $v_\mathrm{Ego}$ frequently exceeds a speed requirement of $v_\mathrm{sup}$, with the duration and the extent of the deviation depending on the control system. Finally, instead of physical limits, we assume an emergency deceleration of \mbox{$a_\mathrm{max} = \qty{6.5}{\metre\per\second\squared}$}. This represents the lower end of the achievable mean deceleration of conventional passenger cars, taking into account possible friction properties~\cite{vangi_evaluation_2007}. Friction is not fully known during vehicle operation and is particularly affected by weather conditions, road surface, and tire wear.

\begin{figure}[t]
\centering
\includegraphics[width=\columnwidth]{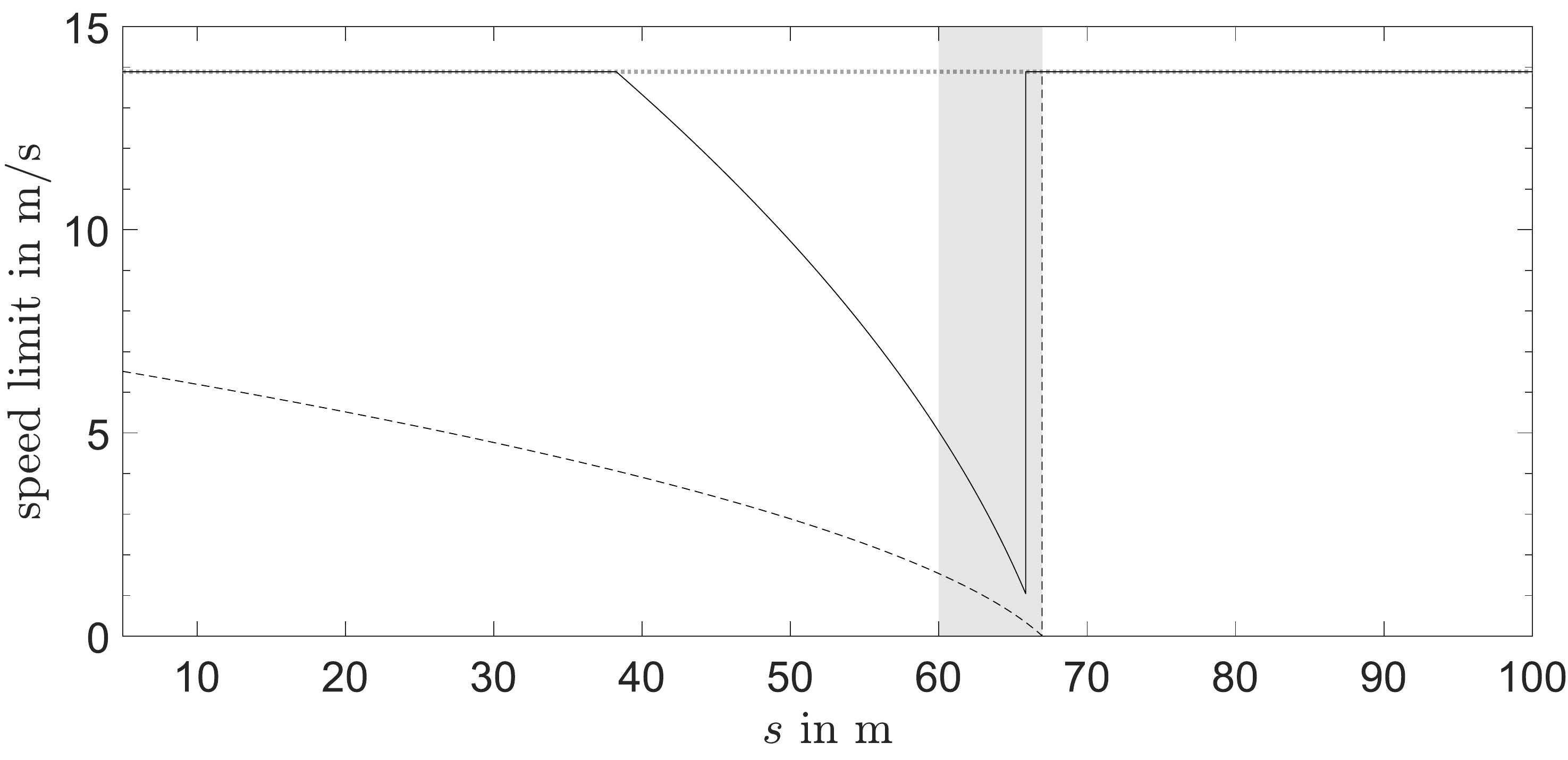}
\caption{Speed limit profile of the ego-vehicle while passing a large parked vehicle with a lateral distance of \qty{1}{\metre} with the introduced example assumptions (solid) and extremely pessimistic assumptions (dashed). The highlighted area indicates the position of the parked vehicle next to the ego-lane and the dotted gray line indicates the posted speed limit.}
\label{fig:SpeedLimitProfile}
\end{figure}

After specifying example assumptions for pedestrian speed, reaction time, and deceleration capability, we can analyze the situation shown in \cref{fig:IniSituation}, where an automated vehicle is driving on an urban road (speed limit \qty{50}{\kilo\metre\per\hour}) and passes a large recreational vehicle parked next to the lane. We assume that, due to oncoming traffic, the effective lateral distance to the large vehicle parked next to the ego-lane is only \qty{1}{\metre} (this is also the distance used in the Euro NCAP CPNC-50 scenario). The speed limit in the course of the route determined using the presented calculation formulas and the specified parameter values is shown by the solid line in \cref{fig:SpeedLimitProfile}. It can be seen that due to the assumed relatively high default speed on the road section, a reduction of the driving speed based on our dynamic speed limitation becomes necessary already about \qty{20}{\metre} before actually reaching the road segment next to the parked vehicle. We also see that the speed is limited to less than \qty{1}{\metre\per\second} for a short distance. On the other hand, we observe that due to the assumed front sensor, the occluded area already loses its relevance for the speed limitation even before the ego-vehicle has passed the parked vehicle with its front line.

\subsection{Resulting Speed Profile in a Challenging Scenario}
\label{ssec:results}

\begin{figure}[b]
\centering
\includegraphics[scale=0.125]{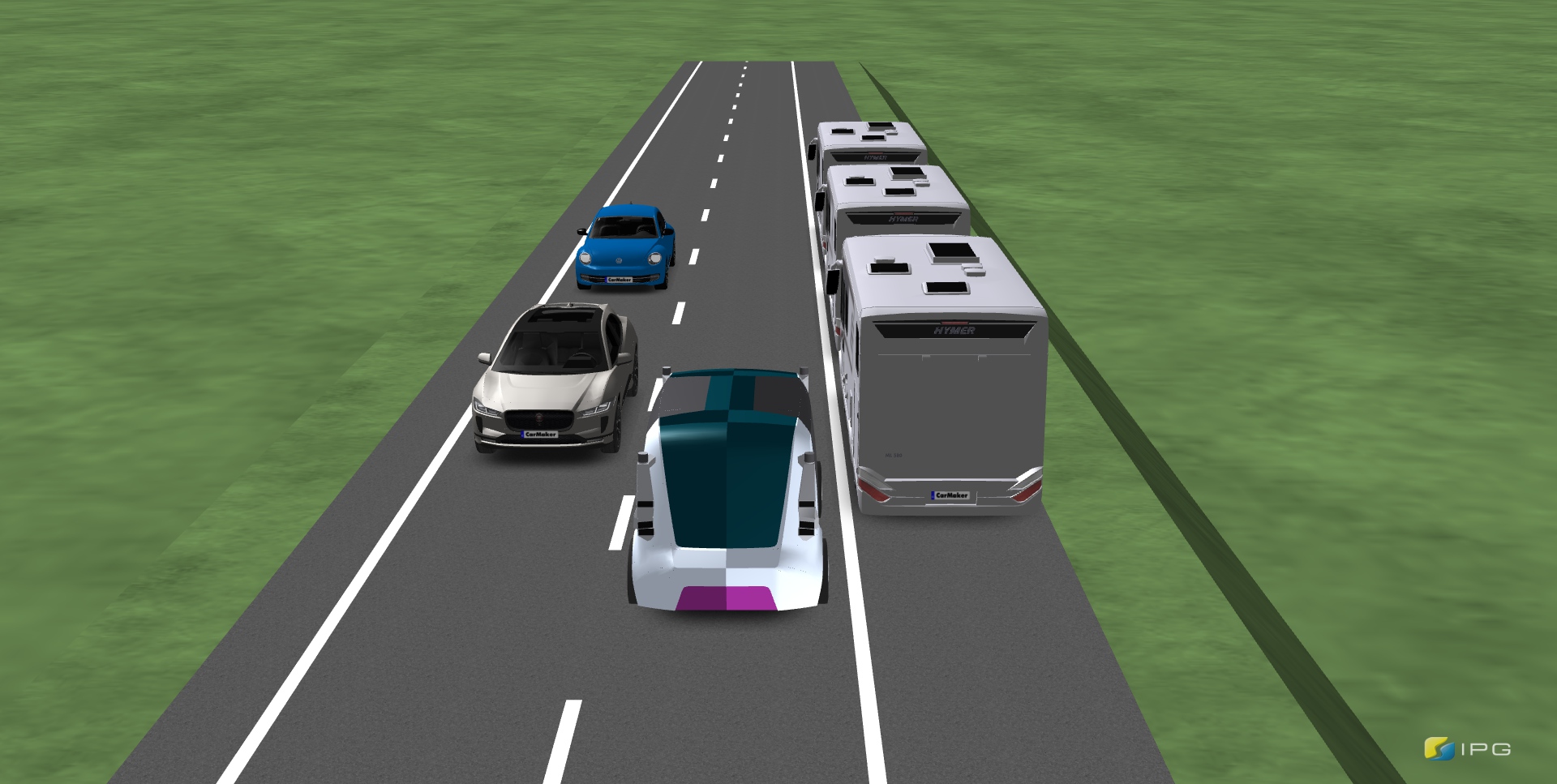}
\caption{Simulated scenario of the UNICAR\emph{agil} concept vehicle passing three critical occlusions in a narrow street}
\label{fig:UNICARSituation}
\end{figure}

\begin{figure}[t]
\centering
\includegraphics[scale=0.3]{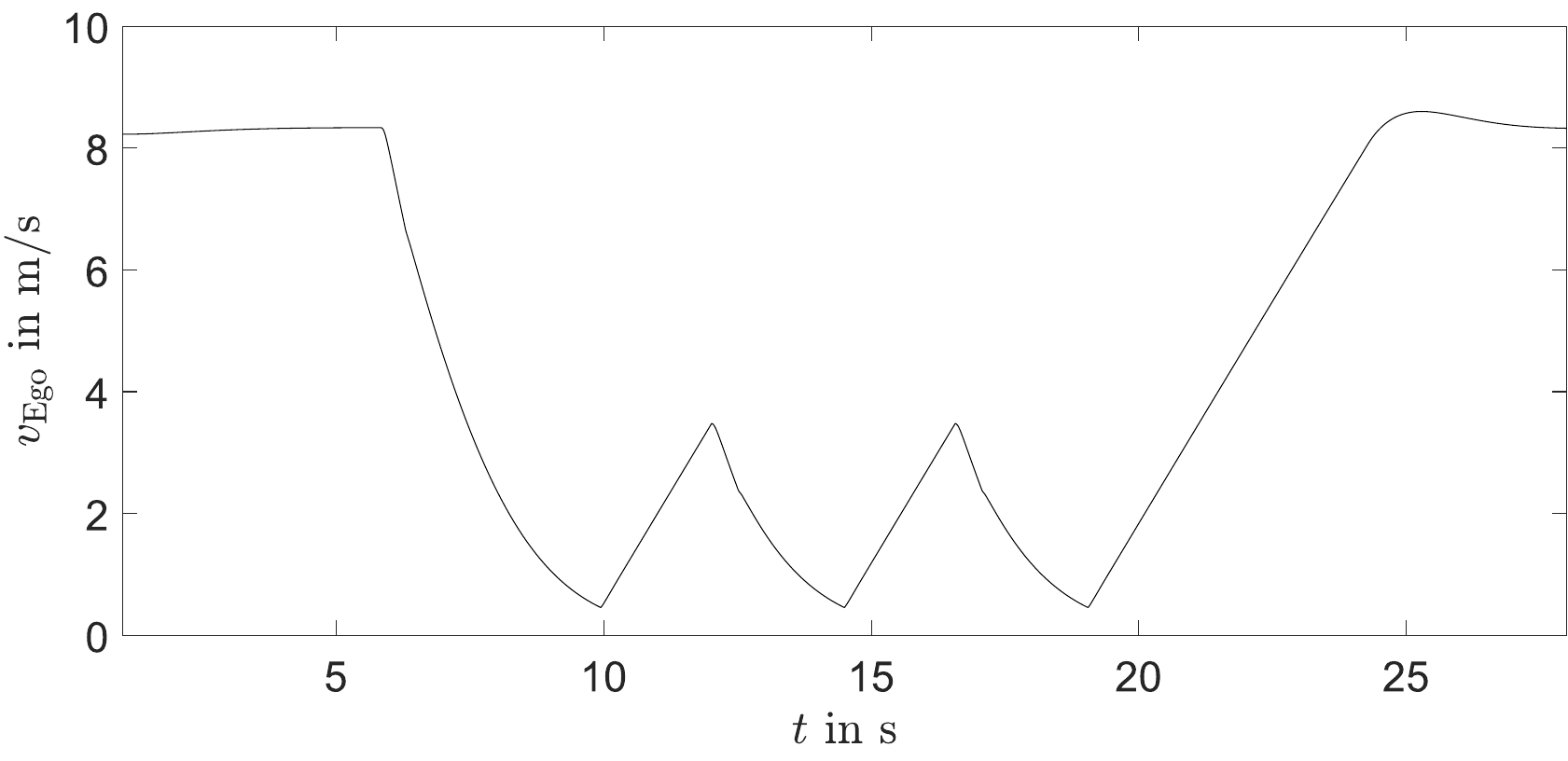}
\caption{Speed profile of the concept vehicle passing three critical occlusions with a lateral distance of only \qty{0.325}{\metre} in simulation}
\label{fig:SimulationSpeedProfile}
\end{figure}

Finally, we use the simulation environment to analyze the actual speed profile when following the calculated speed limit in a challenging scenario. We also investigate the impact on the time it takes to complete the route segment. The initial situation is shown in \cref{fig:UNICARSituation}. We create a scenario in which the UNICAR\emph{agil} concept vehicle has to drive along critical occlusions caused by three recreational vehicles. Due to oncoming traffic, the \qty{2.1}{\metre} wide ego-vehicle is forced to pass the vehicles parked on the side of the road with a distance of only \qty{0.325}{\metre}. We assume that the posted speed limit is \qty{30}{\kilo\metre\per\hour} due to the narrow urban conditions. Otherwise, we adopt all the assumed parameter values from above. The speed profile resulting from the simulation is shown in \cref{fig:SimulationSpeedProfile}.

\cref{fig:SimulationSpeedProfile} indicates that using the example assumptions, the ego-vehicle moves slower than walking speed at each occlusion due to our speed limitation.
\Cref{Sim_results} presents the minimum calculated speed limit, the average speed for the entire route segment, the average speed while passing the parked vehicles, and the total travel time for no occlusion treatment, our example assumptions, and extremely pessimistic assumptions.

\begin{table}[t]
\caption{Summary of Simulation Results}
\label{Sim_results}
\setlength{\tabcolsep}{3pt}
\begin{center}
	\begin{tabular}{lccc}
		\toprule
					& No occlusion				& Example						& Extreme \\
					&treatment					&assumptions					&assumptions\\
		\midrule
		Min. speed limit		& $-$			& \qty{0.46}{\metre\per\second}		& \qty{0.0064}{\metre\per\second}\\
		Avg. speed (route segment) 		& \qty{8.33}{\metre\per\second}		&  \qty{4.82}{\metre\per\second}	& \qty{1.84}{\metre\per\second}\\
		Avg. speed (passing)		& \qty{8.33}{\metre\per\second}	& \qty{2.86}{\metre\per\second}	& \qty{0.52}{\metre\per\second}\\
		Total travel time	& \qty{16.8}{\second}		&\qty{29.1}{\second} 	&\qty{76.0}{\second} \\
		\bottomrule
	\end{tabular}
\end{center}
\end{table}

\section{DISCUSSION}
\label{sec:discussion}

Our simulation results clearly show that the problem of risk due to occluded areas is not solved by a speed limitation model. The calculation formula causes the vehicle to decelerate to a near standstill in an extremely uncomfortable manner when approaching any occluded area caused by parked vehicles. Nevertheless, in our simulation, the risk of collisions is not generally eliminated, as we have used example assumptions that are far from a worst-case scenario. In particular, the assumed parameters of deceleration capability and pedestrian speeds should be mentioned here. In winter conditions, for example, achievable decelerations can be less than \qty{1}{\metre\per\second\squared}, and, as noted above, humans can potentially reach speeds above \qty{10}{\metre\per\second} when running. Moreover, children on bicycles, roller skates, etc. can move even faster.

In addition, our speed limitation model does not take into account other ways that humans can suddenly appear on the roadway. For example, besides the possibility of children emerging from occluded areas between cars, children could also jump out of one of the cars at the side of the road. Consequently, the result of our simulation is not an illustration of a vehicle behavior that avoids every possible collision, but an illustration of the result of an explicit treatment of occluded areas that enables collision avoidance in \emph{some} situations. Thus, if the assumptions made in \Cref{ssec:assump} for our simulative evaluation were made during development, they would create residual risk that would need to be analyzed and evaluated. Where an actual quantification of residual risk would require information on the frequency of pedestrians entering the ego-lane from occlusions within the operational design domain~\cite{on-road_automated_driving_orad_committee_sae_2021,international_organization_for_standardization_iso_2022}. Finally, our analysis excludes the possibility that driving speeds might be perceived as \emph{too fast} by key stakeholders on a purely subjective basis and need to be limited for that reason alone~\cite{stange_is_2022}.

It should be noted, however, that an extremely problematic case has been constructed from the point of view of the ego-vehicle. If smaller vehicles with flat hoods are parked at the side of the road, a roof-mounted environment sensor system could potentially rule out the presence of humans long before the vehicle reaches the occluded area. Nevertheless, for our chosen case with large recreational vehicles, the simulation provides a clear rationale for wide angle front sensors that can detect pedestrians. Otherwise, if only roof-mounted or windshield-mounted sensors were used to detect pedestrians, the speed reduction at small lateral distances would be even more dramatic under our calculation formula.

What we see is a conflict between the expectation that automated vehicles will avoid collisions with pedestrians in principle, and yet operate in the same traffic system as today, with its inherent risk. In this paper, even attempting to avoid many of the collisions caused by pedestrians stepping out of occlusions results in the vehicle moving very slowly and uncomfortably on the road section examined. The flow of traffic on the road would likely be affected unacceptably by the simulated vehicle behavior. Therefore, if the risk of pedestrian collisions is to be significantly reduced by automated vehicles compared to current traffic in urban environments, without requiring road traffic to move very slowly, the risk of people stepping out of occlusions directly adjacent to the roadway should be eliminated (e.g., by adapting the infrastructure). Alternatively, there could be a public consensus that some of these accidents are unavoidable and will remain unavoidable even if road traffic is automated. However, as already explained in \Cref{sec:intro}, the decision that explicit occlusion treatment is not necessary is definitely beyond the competence of the developers and must be made in a public debate~\cite{resnik_precautionary_2023,maurer_wirtschaft_2005}.

\addtolength{\textheight}{-0.372cm}   

\section{CONCLUSIONS}

In this paper, we have demonstrated how assumptions regarding occluded areas in urban environments affect automated vehicle guidance. For this purpose, we introduced a dynamic speed limitation model and evaluated it simulatively based on example assumptions. We concluded that the behavior shown in the simulation does not provide a suitable solution for safe automated driving. Rather, we were able to show that a general problem is to determine to what extent predictive collision avoidance is appropriate and necessary. This question arises in both the development and certification of automated driving systems.

Our work on occlusion treatment is only a simple example that illustrates the urgent need to better communicate design conflicts in the development of automated vehicles to a broad public, in order to create a basis for joint decisions on socially acceptable residual risks. In particular, decisions about which critical scenarios do not need to be handled collision-free should not be made by developers alone. The challenge of finding suitable forms and ways for this communication is one of the key aspects of the recently launched research project AUTOtech.\emph{agil}.




\section*{ACKNOWLEDGMENT}

We thank Kuihong Chen for his assistance in preparing, conducting, and analyzing the simulative evaluation. Furthermore, we thank current and former colleagues for discussions surrounding the presented use case and the speed limitation model. Finally, we thank Sonja Luther for proofreading.


\bibliographystyle{IEEEtran}
\bibliography{IEEEabrv,root}

\begin{thebibliography}{10}
\providecommand{\url}[1]{#1}
\csname url@samestyle\endcsname
\providecommand{\newblock}{\relax}
\providecommand{\bibinfo}[2]{#2}
\providecommand{\BIBentrySTDinterwordspacing}{\spaceskip=0pt\relax}
\providecommand{\BIBentryALTinterwordstretchfactor}{4}
\providecommand{\BIBentryALTinterwordspacing}{\spaceskip=\fontdimen2\font plus
\BIBentryALTinterwordstretchfactor\fontdimen3\font minus
  \fontdimen4\font\relax}
\providecommand{\BIBforeignlanguage}[2]{{%
\expandafter\ifx\csname l@#1\endcsname\relax
\typeout{** WARNING: IEEEtran.bst: No hyphenation pattern has been}%
\typeout{** loaded for the language `#1'. Using the pattern for}%
\typeout{** the default language instead.}%
\else
\language=\csname l@#1\endcsname
\fi
#2}}
\providecommand{\BIBdecl}{\relax}
\BIBdecl

\bibitem{nolte_supporting_2020}
M.~Nolte, I.~Jatzkowski, S.~Ernst, and M.~Maurer, ``Supporting {Safe}
  {Decision} {Making} {Through} {Holistic} {System}-{Level} {Representations}
  \& {Monitoring} {\textendash} {A} {Summary} and {Taxonomy} of
  {Self}-{Representation} {Concepts} for {Automated} {Vehicles},'' 2020, \emph{arXiv:2007.13807}.

\bibitem{salem_risk_2023}
N.~F. Salem, T.~Kirschbaum, M.~Nolte, C.~Lalitsch-Schneider, R.~Graubohm,
  J.~Reich, and M.~Maurer, ``Risk {Management} {Core} {\textendash} {Towards}
  an {Explicit} {Representation} of {Risk} in {Automated} {Driving},'' submitted for publication.

\bibitem{international_organization_for_standardization_isoiec_2014}
  \emph{{Safety} aspects
  {\textemdash} {Guidelines} for their inclusion in standards}, International Organization for Standardization and International Electrotechnical
Commission Standard {ISO}/{IEC} {GUIDE} 51:2014.

\bibitem{european_climate_infrastructure_and_environment_executive_agency_eu_2022}
{European Climate, Infrastructure and Environment Executive Agency}, ``{EU}
  {Road} {Safety}: {Towards} {\textquotedblleft}{Vision}
  {Zero}{\textquotedblright},'' Nov. 2022.

\bibitem{international_organization_for_standardization_iso_2022}
  \emph{{Road} vehicles {\textemdash} {Safety} of the intended functionality}, International Organization for Standardization Standard {ISO} 21448:2022.

\bibitem{resnik_precautionary_2023}
\BIBentryALTinterwordspacing
D.~B. Resnik and S.~L. Andrews, ``\BIBforeignlanguage{en}{A precautionary
  approach to autonomous vehicles},'' \emph{\BIBforeignlanguage{en}{AI Ethics}}, Mar. 2023, doi: {10.1007/s43681-023-00277-6}.
\BIBentrySTDinterwordspacing

\bibitem{maurer_wirtschaft_2005}
\BIBentryALTinterwordspacing
K.~Homann, ``\BIBforeignlanguage{de}{Wirtschaft und gesellschaftliche
  {Akzeptanz}: {Fahrerassistenzsysteme} auf dem {Pr{\"u}fstand}},'' in
  \emph{\BIBforeignlanguage{de}{Fahrerassistenzsysteme mit maschineller
  {Wahrnehmung}}}, M.~Maurer and C.~Stiller, Eds.\hskip 1em plus 0.5em minus
  0.4em\relax Berlin/Heidelberg, Germany: Springer-Verlag, 2005, pp. 239--244, doi: {10.1007/3-540-27137-6\_11}.
\BIBentrySTDinterwordspacing

\bibitem{maurer_hochautomatisiertes_2018}
M.~Maurer, ``Hochautomatisiertes und vollautomatisiertes {Fahren},'' in
  \emph{56. {Deutscher} {Verkehrsgerichtstag} 2018},\hskip 1em plus 0.5em minus 0.4em\relax Goslar,
  Germany, 2018, pp. 43--57.

\bibitem{mobileye_implementing_2018}
{Mobileye}, ``Implementing the {RSS} {Model} on {NHTSA} {Pre}-{Crash}
  {Scenarios},'' 2018.

\bibitem{nister_introduction_2019}
D.~Nist{\'e}r, H.-L. Lee, J.~Ng, and Y.~Wang, ``An {Introduction} to the
  {Safety} {Force} {Field},'' NVIDIA Corp., Santa Clara, CA, USA, Mar. 2019.

\bibitem{wang_potential_2022}
D.~Wang, W.~Fu, Q.~Song, and J.~Zhou, ``Potential risk
  assessment for safe driving of autonomous vehicles under occluded vision,''
  \emph{Sci. Rep.}, vol.~12, no. 4981, Dec.
  2022, doi: {10.1038/s41598-022-08810-z}.

\bibitem{jeong_collision_2020}
Y.~Jeong, J.~Yoo, Y.~Yoon, and K.~Yi, ``Collision {Preventive} {Velocity}
  {Planning} based on {Static} {Environment} {Representation} for {Autonomous}
  {Driving} in {Occluded} {Region},'' in \emph{2020 {IEEE} {Intelligent} {Vehicles} {Symp.}},\hskip 1em plus 0.5em minus 0.4em\relax Las
  Vegas, NV, USA, 2020, pp. 425--430, doi: {10.1109/IV47402.2020.9304540}.

\bibitem{nolte_representing_2018}
M.~Nolte, S.~Ernst, J.~Richelmann, and M.~Maurer, ``Representing the {Unknown}
  {\textendash} {Impact} of {Uncertainty} on the {Interaction} between
  {Decision} {Making} and {Trajectory} {Generation},'' in \emph{2018 {IEEE} {Int.} {Conf.} {Intelligent}
  {Transportation} {Systems}},\hskip 1em plus 0.5em minus 0.4em\relax Maui, HI, USA, 2018,
  pp. 2412--2418, doi: {10.1109/ITSC.2018.8569490}.

\bibitem{intelligent_transportation_systems_committee_2846-2022_2022}
  \emph{{Assumptions} in {Safety}-{Related} {Models} for {Automated} {Driving}
  {Systems}}, {IEEE} Standard 2846-2022.


\bibitem{european_new_car_assessment_programme_test_2021}
{European New Car Assessment Programme}, ``Test
  {Protocol} {\textendash} {AEB} {VRU} systems,'' Version 3.0.4, Apr. 2021.

\bibitem{insurance_institute_for_highway_safety_pedestrian_2022}
{Insurance Institute for Highway Safety},
  ``\BIBforeignlanguage{English}{Pedestrian {Autonomous} {Emergency} {Braking}
  {Test} {Protocol}},'' Version III, Aug. 2022.

\bibitem{international_organization_of_motor_vehicle_manufacturers_future_2019}
{International Organization of Motor Vehicle Manufacturers}, ``Future
  {Certification} of {Automated}/{Autonomous} {Driving} {Systems},'' presented at the UNECE Working Party Automated/Auton. Connected Vehicles 2nd session, Geneva,
  Switzerland, 2019, GRVA-02-09.

\bibitem{european_commission_directorate-general_for_internal_market_industry_entrepreneurship_and_smes_commission_2022}
{European Commission, Directorate-General for Internal Market, Industry,
  Entrepreneurship and SMEs}, ``Commission {Implementing} {Regulation} ({EU})
  2022/1426 of 5 {August} 2022 laying down rules for the application of
  {Regulation} ({EU}) 2019/2144 of the {European} {Parliament} and of the
  {Council} as regards uniform procedures and technical specifications for the
  type-approval of the automated driving system ({ADS}) of fully automated
  vehicles,'' \emph{Off. J. Eur. Union}, vol.~65, no. L221,
  pp. 1--64, Aug. 2022.
  
\bibitem{schepens_mechanics_1998}
\BIBentryALTinterwordspacing
B.~Schepens, P.~A. Willems, and G.~A. Cavagna, ``The
  mechanics of running in children,'' \emph{J. Physiol.}, vol. 509, no.~3, pp. 927--940, Jun. 1998, doi: {10.1111/j.1469-7793.1998.927bm.x}.
\BIBentrySTDinterwordspacing

\bibitem{european_commission_directorate_general_for_research_and_innovation_ethics_2020}
\BIBentryALTinterwordspacing
{European Commission, Directorate-General for Research and Innovation},
  ``Ethics of connected and automated vehicles:
  recommendations on road safety, privacy, fairness, explainability and
  responsibility,'' Publications Office, Luxembourg, Luxembourg, 2020, doi: {10.2777/035239}.
\BIBentrySTDinterwordspacing

\bibitem{feldmanis_principle_2019}
L.~Feldmanis, ``The {Principle} of {Trust} for {Exceptions} to the
  {Non}-{Regression} {Clause} in the {Case} of {Delict} of {Negligence},''
  \emph{Juridica Int.}, vol.~28, pp. 86--94, Nov. 2019, doi: {10.12697/JI.2019.28.10}.

\bibitem{frey_psychologische_2018}
A.~Frey and T.~M. Gasser, ``\BIBforeignlanguage{german}{Psychologische und rechtliche {Aspekte} geltender
  {\quotedblbase}{Bewegungsregelung}{\textquotedblleft} im
  {Stra{\ss}enverkehr}},'' \BIBforeignlanguage{german}{Oberseminar Elektronische Fahrzeugsysteme, TU Braunschweig}, Feb. 2018.

\bibitem{international_organization_for_standardization_isotr_2020}
{International Organization for Standardization},
  ``{Road} vehicles
  {\textemdash} {Safety} and cybersecurity for automated driving systems
  {\textemdash} {Design}, verification and validation,''\hskip 1em plus 0.5em
  minus 0.4em\relax Geneva, Switzerland, {ISO}/{TR}~4804:2020.

\bibitem{national_science__technology_council_ensuring_2020}
{National Science \& Technology Council} and {United States Department of
  Transportation}, ``\BIBforeignlanguage{English}{Ensuring {American}
  {Leadership} in {Automated} {Vehicle} {Technologies} {\textendash}
  {Automated} {Vehicles} 4.0},'' Jan. 2020.

\bibitem{national_highway_traffic_safety_administration._federal_2016}
\BIBentryALTinterwordspacing
{National Highway Traffic Safety Administration.},
  ``\BIBforeignlanguage{English}{Federal {Automated} {Vehicles} {Policy}
  {\textendash} {Accelerating} the {Next} {Revolution} {In} {Roadway}
  {Safety}},'' Sep. 2016.
\BIBentrySTDinterwordspacing

\bibitem{shalev-shwartz_formal_2017}
\BIBentryALTinterwordspacing
S.~Shalev-Shwartz, S.~Shammah, and A.~Shashua, ``On a {Formal} {Model} of
  {Safe} and {Scalable} {Self}-driving {Cars},'' 2018, \emph{arXiv:1708.06374v6}.
\BIBentrySTDinterwordspacing

\bibitem{naumann_safe_2019}
M.~Naumann, H.~Konigshof, M.~Lauer, and C.~Stiller, ``Safe but not
  {Overcautious} {Motion} {Planning} under {Occlusions} and {Limited} {Sensor}
  {Range},'' in \emph{2019 {IEEE} {Intelligent} {Vehicles} {Symp.}},\hskip 1em plus 0.5em minus 0.4em\relax Paris, France,
  2019, pp. 140--145, doi: {10.1109/IVS.2019.8814251}.

\bibitem{yu_occlusion-aware_2019}
M.-Y. Yu, R.~Vasudevan, and M.~Johnson-Roberson, ``Occlusion-{Aware} {Risk}
  {Assessment} for {Autonomous} {Driving} in {Urban} {Environments},''
  \emph{IEEE Robot. Automat. Lett.}, vol.~4, no.~2, pp. 2235--2241,
  Apr. 2019, doi: {10.1109/LRA.2019.2900453}.

\bibitem{bouton_scalable_2018}
M.~Bouton, A.~Nakhaei, K.~Fujimura, and M.~J. Kochenderfer, ``Scalable
  {Decision} {Making} with {Sensor} {Occlusions} for {Autonomous} {Driving},''
  in \emph{2018 {IEEE} {Int.} {Conf.} {Robotics}
  {Automation}},\hskip 1em plus 0.5em minus 0.4em\relax Brisbane, Australia, 2018, pp. 2076--2081, doi: {10.1109/ICRA.2018.8460914}.

\bibitem{schratter_pedestrian_2019}
M.~Schratter, M.~Bouton, M.~J. Kochenderfer, and D.~Watzenig, ``Pedestrian
  {Collision} {Avoidance} {System} for {Scenarios} with {Occlusions},'' in
  \emph{2019 {IEEE} {Intelligent} {Vehicles} {Symp.}},\hskip 1em
  plus 0.5em minus 0.4em\relax Paris, France, 2019, pp. 1054--1060, doi: {10.1109/IVS.2019.8814076}.

\bibitem{woopen_unicaragil_2018}
T.~Woopen \emph{et al.}, ``{UNICARagil} -
  {Disruptive} {Modular} {Architectures} for {Agile}, {Automated} {Vehicle}
  {Concepts},'' in \emph{27th {Aachen} {Colloq.}
  {Automobile} {Engine} {Technology} 2018},\hskip 1em plus 0.5em minus
  0.4em\relax Aachen, Germany, 2018, pp. 663--694, doi:
  {10.18154/RWTH-2018-229909}.

\bibitem{international_organization_for_standardization_iso_2010}
 \emph{{Safety} of machinery
  {\textemdash} {Positioning} of safeguards with respect to the approach speeds
  of parts of the human body}, International Organization for Standardization Standard {ISO} 13855:2010.

\bibitem{rieken_lidar-based_2020}
\BIBentryALTinterwordspacing
J.~Rieken and M.~Maurer, ``A {LiDAR}-based real-time capable {3D} {Perception}
  {System} for {Automated} {Driving} in {Urban} {Domains},'' 2020, \emph{arXiv:2005.03404v1}.
\BIBentrySTDinterwordspacing

\bibitem{vangi_evaluation_2007}
\BIBentryALTinterwordspacing
D.~Vangi and A.~Virga, ``Evaluation of emergency
  braking deceleration for accident reconstruction,''
  \emph{Veh. Syst. Dyn.}, vol.~45, no.~10, pp.
  895--910, Oct. 2007, doi: {10.1080/00423110701538320}.
\BIBentrySTDinterwordspacing

\bibitem{on-road_automated_driving_orad_committee_sae_2021}
\emph{{Taxonomy} and {Definitions} for {Terms} {Related} to {Driving}
  {Automation} {Systems} for {On}-{Road} {Motor} {Vehicles}}, SAE Standard J3016\_202104.

\bibitem{stange_is_2022}
\BIBentryALTinterwordspacing
V.~Stange, M.~Steimle, M.~Maurer, and M.~Vollrath, ``Is
  the automated vehicle {\textquotedblleft}aware{\textquotedblright} of the
  pedestrian? {Examining} driving behavior adaptation as a cue to inform the
  passenger of a potential hazard,''
  \emph{Transp. Res. Interdiscip. Perspect.}, vol.~16, 100701, Dec. 2022, doi: {10.1016/j.trip.2022.100701}.
\BIBentrySTDinterwordspacing

\end{thebibliography}

\end{document}